\documentclass{PoS}
\usepackage{gensymb}
\usepackage{subcaption}
\RequirePackage{lineno}
\usepackage{amsmath}
\usepackage[percent]{overpic}
\usepackage{graphics}
\usepackage{graphicx}

\title{Correlated GeV--TeV Gamma-Ray Emission from Extended Sources in the Cygnus Region}

\ShortTitle{Cygnus Cocoon}

\author{\speaker{B. Hona}\\
        Michigan Technological University\\
        E-mail: \email{bhona@mtu.edu}}

\author{{A. Robare}\\
        Michigan Technological University\\
        E-mail: \email{alrobare@mtu.edu}}
        
\author{{H. Fleischhack}\\
        Michigan Technological University\\
        E-mail: \email{hfleisch@mtu.edu}}
        
\author{{ P. Huentemeyer}\\
        Michigan Technological University\\
        E-mail: \email{petra@mtu.edu}}   
        
\author{{for the HAWC Collaboration}\\
For a complete author list see www.hawc-observatory.org/collaboration/icrc2017.php}

\abstract{The Cygnus arm of our galaxy is a source-rich and complex region hosting multiple gamma-ray source types such as pulsar wind nebulae (PWN), supernova remnants, binary systems, and star clusters. The High Altitude Water Cherenkov (HAWC) observatory has been collecting data continuously since 2015 and has reported five sources within the Cygnus region. Several other instruments have also observed gamma-ray sources in this region. For instance, Fermi-LAT found gamma-ray emission at GeV energies due to a $“Cocoon”$ of freshly accelerated cosmic rays, which is co-located with a known PWN TeV 2032+4130 seen by several TeV gamma-ray observatories. TeV J2032+4130 is likely powered by the pulsar PSR J2032+4127 based on the multi-wavelength observation and asymmetric morphology reported by VERITAS. The study of HAWC data will provide more information regarding the morphology, emission origin, and the correlation with the GeV emission. This presentation will discuss the analysis of data collected with the HAWC instrument and the Fermi-LAT and the results obtained to provide a deeper understanding of the Cygnus Cocoon across five decades of energy range.}

\FullConference{35th International Cosmic Ray Conference --- ICRC2017\\
		10--20 July, 2017\\
		Bexco, Busan, Korea}

\begin{document}

\section{Introduction}
In our Galaxy,  cosmic rays can be accelerated up to a few PeV. Investigating
the physics behind the  production of very high energy gamma-ray emission is a pivotal to understanding of
the cosmic-ray acceleration in the Galaxy. The Cygnus region (70\degree<l<85\degree and -4\degree<b<4\degree)  is akin to a laboratory providing a number of accelerator sites with multiple gamma-ray sources to study. A particularly interesting site is the
Cocoon region observed by Fermi-LAT which consists of "a Cocoon of freshly accelerated cosmic
rays" and is responsible for extended hard GeV emission seen at this location \cite{Acker11}.

The Cocoon, so far, has no identified counterpart in other wavelengths. Observations in (2--10) keV by the Suzaku observatory concluded that extended X-ray emission detected at the location after subtraction of point sources, small-scale structures from X-ray images and Cosmic X-ray background is related to Galactic ridge X-ray emission rather than the Cocoon \cite{suzaku}. In the TeV range, three sources have spectra consistent with the extrapolation of the Fermi spectrum at higher energies and might be related to the GeV Cocoon \cite{abdo12, Bartoli14, Abey17}.  The spectral comparison alone is not sufficient to establish the relationship between the GeV and TeV emission. This analysis focuses on morphological studies at the Cocoon region to understand the origin of the TeV gamma-ray emission and the correlation between the GeV--TeV emission.

\section{Fermi Cocoon}
Fermi-LAT is a satellite-based gamma ray observatory, sensitive to gamma-ray emission in the 20 MeV--300 GeV range \footnote{\texttt{https://fermi.gsfc.nasa.gov/science/instruments/table1-1.html}}. Fermi-LAT observation of the Cygnus region revealed an extended excess after subtraction  known point sources, extended emission from Gamma Cygni and background using a diffuse emission model specifically developed for the Cygnus region in the energy range of 1 GeV -100 GeV \cite{Acker11, fermi2}. This excess emission is detected with $10.1\sigma $ significance above 1 GeV and is best described by a Gaussian width of $\sigma = 2.0\degree \pm 0.2\degree $. Observations of $8 \mu m$ map shows that the gamma-ray excess is surrounded by regions with high infrared luminosity. The emission morphology is similar to a cavity bounded by ionization fronts which is formed due to the stellar winds of massive star clusters and hence, the GeV emission is called a Cocoon. The hard spectrum observed by Fermi-LAT for this Cocoon indicates the source of the emission to be freshly-accelerated cosmic rays (CRs). The Cocoon is 50 pc wide and lies between the supernova remnant (SNR) Gamma Cygni and the star cluster Cygnus OB2 \cite{Acker11}. The origin of the cosmic rays in the Cocoon can possibly be attributed to one or both of these objects. 

The Fermi-LAT observatory allows analysis of its data via the publicly available Fermi Science tools  \footnote{\texttt{https://fermi.gsfc.nasa.gov/ssc/data/analysis/}} and Fermi data \footnote{\texttt{https://fermi.gsfc.nasa.gov/cgi-bin/ssc/LAT/LATDataQuery.cgi}}. They were utilised to look into the Cocoon region for 6 additional years of Fermi data. The residual count map shown in Fig.\ref{fig:cocoon} was obtained after subtracting all known point sources, Gamma Cygni, and publicly available background models. The map shows an extended emission at the location published in the Fermi paper. The detected emission has a hard spectrum comparable to the spectrum reported by Fermi-LAT. This 8-year Cocoon spectrum is shown as blue points in Fig.\ref{fig:spectra}  and it appears to be consistent with extrapolation of  gamma-ray spectrum obtained by the wide-field TeV observatory to lower energies as discussed in more detail in section 3.3 below.

\section{Cygnus Cocoon Region with HAWC data}
\subsection{The HAWC Observatory}
The High Altitude Water Cherenkov (HAWC) observatory is a wide-field TeV gamma-ray observatory  located at Sierra Negra, Mexico at an altitude of 4100m \cite{Abey13a}. The HAWC gamma-ray instrument  comprises 300 water cherenkov detectors (WCD) in an array and is sensitive to gamma rays in the energy range of 100 GeV to 100 TeV.  Each WCD has four photomultipliers tubes (PMT) at the bottom which detect the Cherenkov light  produced by charged particles from the air showers travelling in the WCD. Air shower events recorded by the detector are reconstructed to extract shower properties such as the direction of the primary particle and the size of the shower \cite{Abey17crab}.

The HAWC data is divided into 9 size bins according to the fraction $f_{hit}$ of PMTs hits which are used for the event reconstruction. Higher bins correspond to higher energy gamma rays and have better angular resolution. The angular resolution (68\% containment radius) for bin 9 is $0.17\degree$ and $1.03\degree$ for bin 1 \cite{Abey17crab}. After the reconstruction process, event and background maps are generated. The background for each bin is estimated using the direct integration method \cite{Abey17crab}. The hadronic cosmic rays that pass gamma/hadron separation cuts during reconstruction form the main background in the analysis of gamma-ray sources. To calculate the statistical significance of the excess against the background, a maximum likelihood  framework is used. 
\subsection{Maximum Likelihood Fit}
The maximum likelihood fit assumes a source model and estimates the free parameters of the model. For a point source model, the model parameters are position and spectrum of a source. In case of an extended source model, apart from these parameters, there are one or more additional parameters depending on the mophology of the model. For the study presented here, the source spectra are characterised by a power-law spectrum given by
\begin{equation}
\dfrac{dN}{dE}=I_{0}\dfrac{E}{E_{0}}^{-\Gamma}
\end{equation}
where $I_{0}$ is the differential flux normalization, $\Gamma$ is the spectral index, and  $E_{0}$ is the pivot energy. 

The parameter values are estimated so as to maximise the likelihood of a model. To calculate goodness of fit between two models, a likelihood ratio test is used:

\begin{equation}
 -2ln\dfrac{L_{0}}{L_{1}} = Test Statistics(TS)
\end{equation}
where $L_0$ is the likelihood for the background-only hypothesis, $ L_1$ is the source-model hypothesis consisting of the source and background for the same data. The likelihood ratio is defined as test statistics (TS) \cite{pat}. According to Wilks' theorem for nested models when the background only hypothesis is true, the TS values follow a $\chi ^2$ distribution with the number of degrees of freedom equal to the difference in the number of free parameters between the hypotheses \cite{wilk}. For example: Given a background only hypothesis (there is no source) and a source-model hypothesis assuming there is a source with flux normalisation N, the difference in the number of free parameters between the hypotheses is 1 (flux normalisation). If the background only hypothesis is true, TS is distributed as $\chi ^2 (Dof = 1)$. Then, $\sqrt{TS}$ can be written as significance of excess of events in units of Gaussian sigma \cite{pat}.
\begin{equation}
\sigma=\sqrt{TS}
\end{equation}
\subsection{TeV Sources in the Cygnus Cocoon Region}
\begin{figure}[b]
\begin{subfigure}{.5\textwidth}
\includegraphics[scale=0.275]{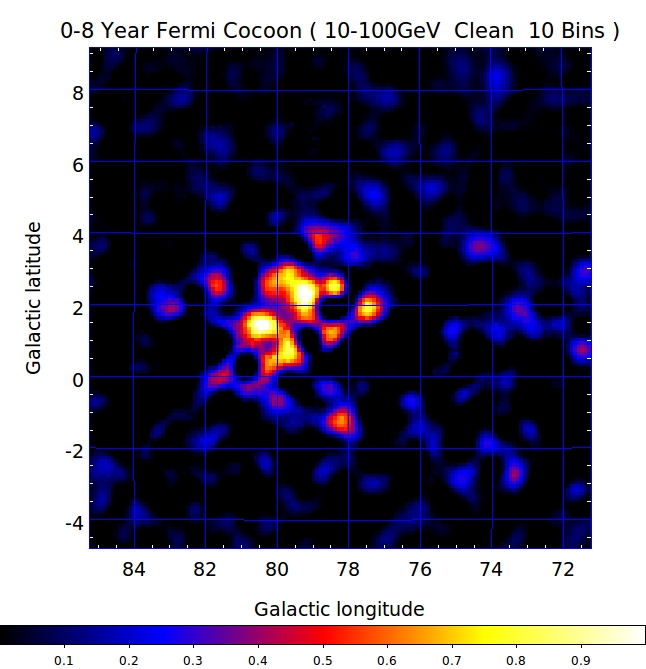}
\caption{Resicual count map of the Cocoon with the\\ pass8 clean data class in Fermi Science tools}
\label{fig:cocoon}
\end{subfigure}
\begin{subfigure}{.5\textwidth}
\begin{overpic}[width=8cm,tics=8, trim={0 3pt  0 42pt}, clip]{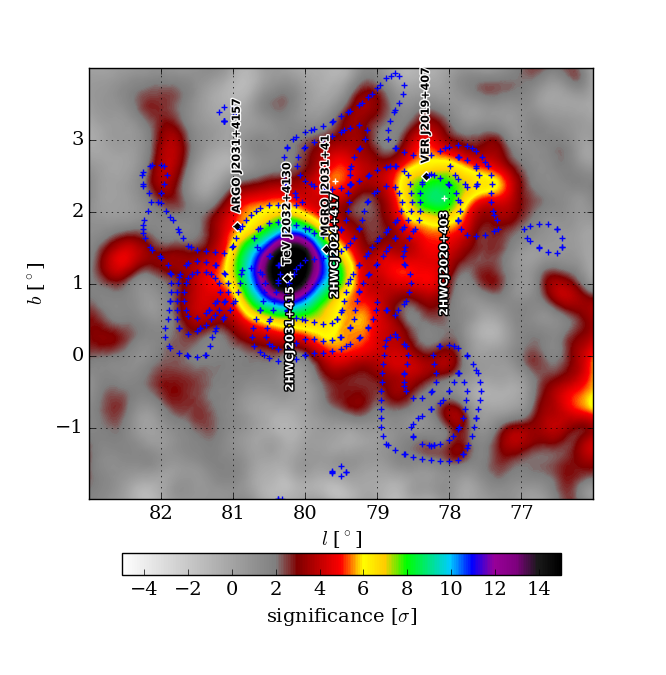}
\put(20,30){\color{white} Preliminary }
\end{overpic}
\caption{Cygnus Cocoon with 25 months of HAWC data overlaid with Fermi Cocoon contour from \cite{Acker11} in blue. Black squares correspond to TeVcat sources, white plus-signs to HAWC sources.}
\label{fig:cocoon1}
\end{subfigure}
\end{figure}

Shown in Fig.\ref{fig:cocoon1} is the significance map of the Cocoon region with 760 days of HAWC data using a point source search  with an index of -2.7. According to the 2HWC catalog \cite{Abey17}, five sources were detected in the Cygnus region, three of which lie in the vicinity of the Cocoon as listed in Table \ref{tab:hawcsources}.
 
\begin{table}[b]
\begin{center}
\caption{HAWC Sources in the Cocoon Region}
\label{tab:hawcsources}
\begin{tabular}{c c c c c} 

\textbf{Name}            &   \textbf{RA }[deg]        &  \textbf{Dec} [deg]      &   \textbf{Nearest TeVCat Source}\\      

2HWC J2020+403      &          305.16           &        40.37          &        VER J2019+407   \\  

2HWC J2024+417      &          306.04           &        41.76          &        MGRO J2031+41 \\ 

2HWC J2031+415      &          307.93           &        41.51           &        TeV J2032+4130 \\
\end{tabular}
\end{center}
\end{table}

The HAWC observatory detects the strongest emission at the \textbf{2HWC J2031+415} location which lies about $0.08\degree$ from the PWN TeV J2032+4130. This PWN was first detected by HEGRA as an unidentified TeV source and has been shown to favor a PWN model associated with a binary pulsar - PSR J2032+4127 by  VERITAS \cite{aha02, aliu}. Properties of the pulsar are as shown in Table \ref{tab:pulsars}  \cite{atnf} \footnote{\texttt{http://www.atnf.csiro.au/people/pulsar/psrcat}}. 

\begin{table}[b]
\begin{center}
\caption{Characteristics of the pulsars. }
\label{tab:pulsars}
\begin{tabular}{c c c c } 
\textbf{Name}            &   \textbf{Age }[Kyr]  &  \textbf{Distance} [pc]   &   $\dot{\textbf{E}}$ [erg/s] \\      
PSR J2032+4127     &          181           &                 1700         &                  1.7e35           \\  
PSR J0633+1746     &          342           &                   250         &                  3.2e34           \\
PSR B0656+14        &           111           &                   288         &                  3.8e34           \\
\end{tabular}
\end{center}
\end{table}

\begin{figure}[b]
\begin{center}
\includegraphics[scale=0.43]{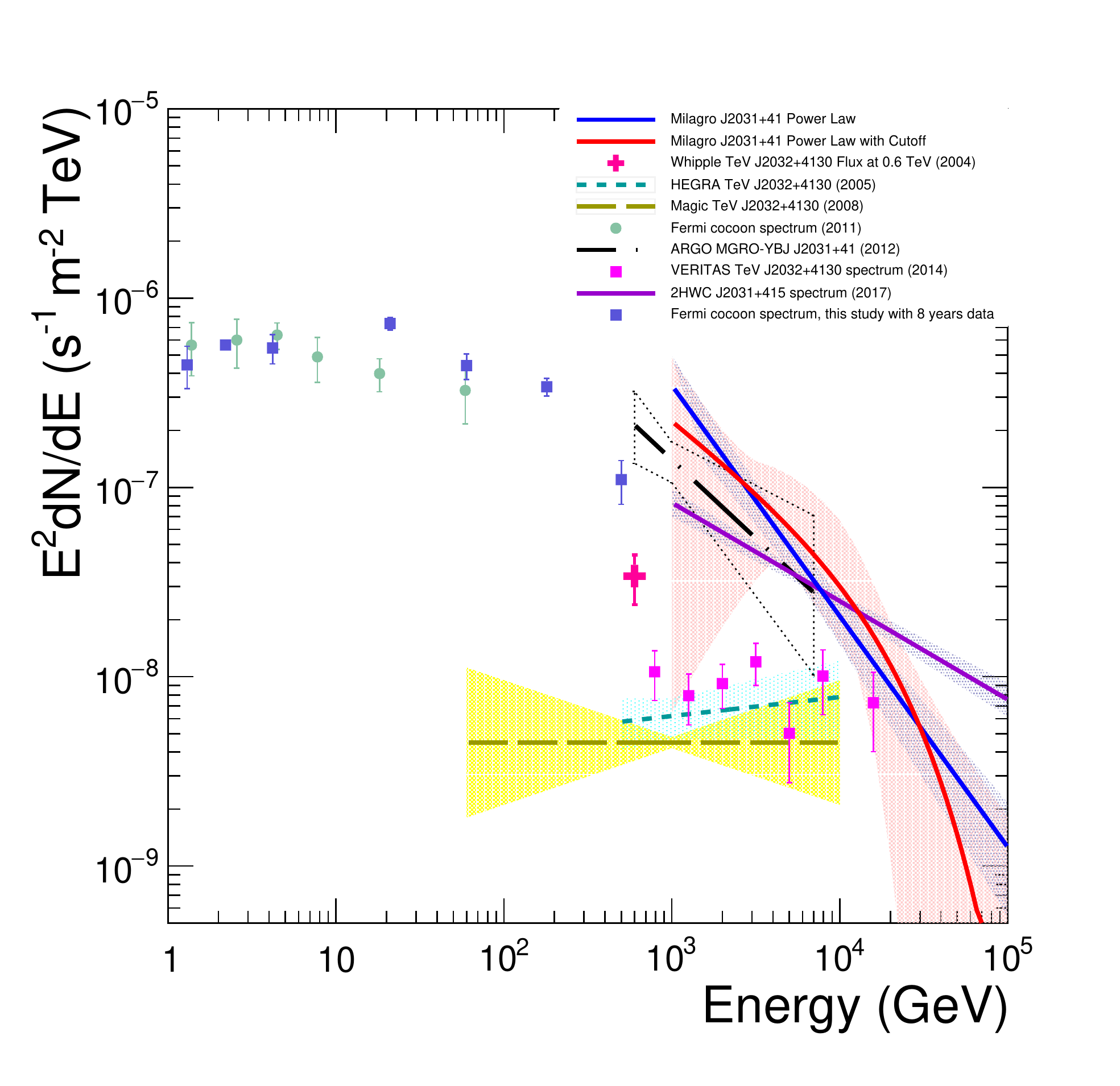}
\caption{Energy spectrum of the Cocoon region as measured by various instruments}
\label{fig:spectra}
\end{center}
\end{figure}

The second strongest emission at the Cocoon region is detected at \textbf{2HWC J2020+403} in the vicinity of the SNR Gamma Cygni. This HAWC source also has no identified counterpart; a possible association could be PSR J2021+4026 or SNR G78.2+2.1 \cite{Abey17}. In the multi-source fit analysis presented in section 4, emission from this location has been modeled as a point source. The HAWC observatory also detects a third TeV source, \textbf{2HWC J2024+417} which is $0.35\degree$ from the Fermi-LAT source 3FGL J2023.5+4126. However, this HAWC source could be a part of the extended morphology of 2HWC J2031+415  \cite{Abey17}. A spectral comparison of 2HWC J2031+415 along with the TeV sources seen by various gamma-ray instruments and the GeV Cocoon  seen by Fermi-LAT is given in Fig.\ref{fig:spectra}. 
The flux measured for 2HWC J2031+415 is closer to the fluxes reported by the Milagro and ARGO experiments. The Milagro and ARGO sources are larger in angular size and have higher fluxes than the fluxes measured by VERITAS and other IACTs \cite{ abdo12, Abey17}.  As shown in Fig.\ref{fig:spectra}, the extrapolation of the HAWC spectrum to lowr GeV energies seems consistent with the Fermi Cocoon spectrum. The same is true for the MGRO J2031+41 spectrum and the ARGO J2031+4157 spectrum \cite{hui13}. ARGO J2031+4157 has been suggested as a counterpart of the Cocoon at TeV energies after subtraction of the PWN emission measured by other instruments \cite{Bartoli14}. The relation between the Cocoon and TeV sources overlapping the region is still unclear and needs further morphological studies.
In order to understand the TeV emission and morphology in the Cocoon region,  various models were tested in the region to disentangle the sources. Preliminary tests in the region indicated the possibility of an extended emission at the 2HWC J2031+415 location, whearas a point source model was used to describe the 2HWC J2020+403 location. The multi-source fit allowed us to look at the Cocoon region after constraining the emission from 2HWC J2031+415 and 2HWC J2020+403. The Multi-Mission Maximum Likelihood (3ML) \cite{Via} \footnote{\texttt{https://github.com/giacomov/3ML}}  software was used for the fit. 

\section{Morphological Studies}
The steps for the studies include selecting a model, deciding the free and fixed parameters, obtaining the fit results and test statistics, and based on the results, making a significance map of the model and the residual.
\subsection{Multi-Source Fit with a Simple Gaussian and a Point Source Model }
\begin{figure}[b]
\begin{center}
\includegraphics[scale=0.55]{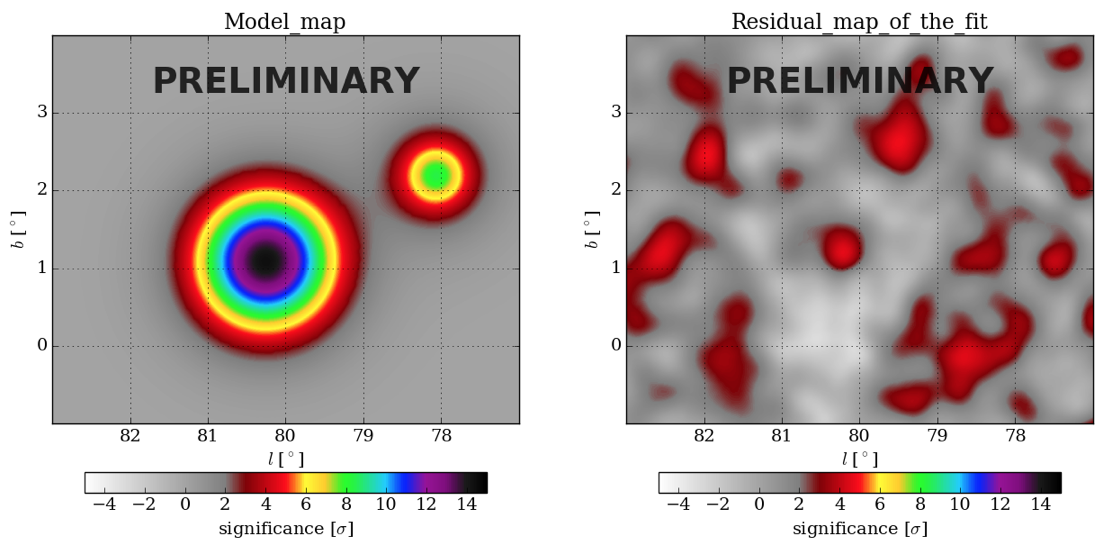}
\caption{ A Simple Gaussian + Point Source Model Fit }
\label{fig:fit1}
\end{center}
\end{figure}

A simple Gaussian shape for the 2HWC J2031+415 was tested. The positions were fixed according to the 2HWC catalog  \cite{Abey17}. The parameters fitted for the Gaussian shape were spectral index, flux normalization and width of the Gaussian. The parameters fitted for the point source were spectral index and flux normalization. Fig.\ref{fig:fit1} shows the model map and the residual after subtracting the model expectations. Multiple significant hotspots close to $5\sigma$ remain along with oversubtraction near the centre in the residual map which indicate that a Gaussian shape is not a good description of the TeV emission in this region. Morphological models based on the physics processes present in this complicated region need to be explored.

\subsection{Multi-Source Fit with a Pulsar Diffusion Model and a Point Source Model }
Since 2HWC J2031+415  overlaps with the PWN, a pulsar diffusion morphology was selected at the location which was developed in the study of  PSR J0633+1746 and PSR B0656+14  \cite{geminga, hao}. This pulsar diffusion model assumes  isotropic diffusion with continuous injection of electrons/positrons with  power-law spectra into the PWN  \cite{atoyan,hao, geminga}. The cooling time of the particles diffusing from the pulsars PSR J0633+1746 and PSR B0656+14  was calculated to be about $10^{4}$ years according to the tested model, which  is lower than the age of the pulsars. The characteristic age of  PSR J2032+4127  is comparable to those of PSR J0633+1746 and PSR B0656+14 as seen in Table \ref{tab:pulsars} \cite{atnf} and is longer than the cooling time for the diffusing particles. Due to this reason, the pulsar diffusion morphology was tested for the 2HWC J2031+415 location.

\begin{figure}[b]
\begin{center}
\includegraphics[scale=0.55]{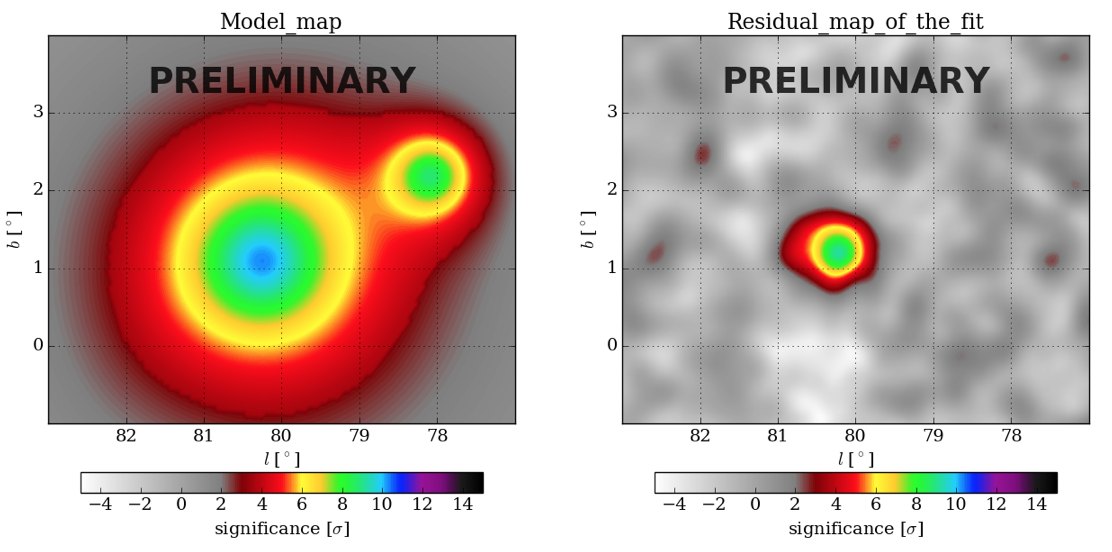}
\caption{A Pulsar Diffusion + Point Source Model Fit }
\label{fig:fit2}
\end{center}
\end{figure}

The free parameters  for this model were spectral index and flux normalization of the two sources and the radius of the diffusion sphere. The residual map  in Fig.\ref{fig:fit2} shows that the tested model accounted for the multiple hotspots that were present while modeling the region as a simple Gaussian. It also improved the TS significantly in comparison to the simple Gaussian model tested at the region.  However, a significant excess of $8\sigma$ remain at the location of $RA=307.6\degree$ and $Dec=41.57\degree$. This remaining flux is about 3.5 \% of the total  flux detected at 2HWC J2031+415 location. The residual map of the fit indicates that the tested model was not a complete description of TeV emission from the Cygnus Cocoon region.  
\section{Conclusion and Discussion}
Based on the results of the tested pulsar diffusion model, it is possible that there is a source overlapping the PWN location, which is  responsible for the remaining TeV emission detected after fitting the emission from the PWN and 2HWC J2020+403. Also, the current model does not include Galactic diffuse emission which might contribute to the TeV emission detected in the Cocoon region. Another possibility is that the tested pulsar diffusion morphology is not a good description of the region as it does not address the processes such as binary nature of the pulsar. Multiple sources close to each other and the binary nature of PSR J2032+4127 pose a challenge in this region for morphological studies. While preliminary steps have been undertaken to disentagle and understand the TeV morphology, it is still work in progress. A morphological study based on the physics processes in the region which also include the GeV Cocoon will be helpful to better understand the gamma-ray emission in the region. For the study presented here, the Fermi-LAT  and HAWC data were investigated separately. A joint analysis with the HAWC and Fermi-LAT data using 3ML is in progress. For the future analysis, HAWC is developing improved energy estimators, which will improve the sensitivity to different spectral shapes \cite{sam}.

\section{Acknowledgement}
We acknowledge the support from: the US National Science Foundation (NSF); the US Department of Energy Office of High-Energy Physics; the Laboratory Directed Research and Development (LDRD) program of Los Alamos National Laboratory; Consejo Nacional de Ciencia y Tecnolog\'{\i}a (CONACyT), Mexico (grants 271051, 232656, 167281, 260378, 179588, 239762, 254964, 271737, 258865, 243290); Red HAWC, Mexico; DGAPA-UNAM (grants RG100414, IN111315, IN111716-3, IA102715, 109916); VIEP-BUAP; the University of Wisconsin Alumni Research Foundation; the Institute of Geophysics, Planetary Physics, and Signatures at Los Alamos National Laboratory; Polish Science Centre grant DEC-2014/13/B/ST9/945.

\end{document}